\documentclass{article}
\usepackage{chngcntr}
\counterwithin{figure}{section}

\usepackage{blindtext} 

\usepackage{microtype} 

\usepackage{subfigmat}
\usepackage{amsmath}
\usepackage{cleveref}
\usepackage{natbib}
\usepackage[english]{babel} 

\usepackage[hmarginratio=1:1,top=32mm,columnsep=10pt]{geometry} 
\usepackage{booktabs} 

\usepackage{lettrine} 

\usepackage{enumitem} 
\usepackage{grffile} 
\setlist[itemize]{noitemsep} 

\usepackage{abstract} 

\usepackage{titlesec} 
\titleformat{\section}[block]{\large\scshape\centering}{\thesection.}{1em}{} 
\titleformat{\subsection}[block]{\large}{\thesubsection.}{1em}{} 
\usepackage{fancyhdr} 
\usepackage{titling} 

\usepackage{graphicx}

\usepackage{float}

\pretitle{\begin{center}\Huge\bfseries} 
\posttitle{\end{center}} 
\title{A seismic tsunami in the Irish annals, recorded at Iona in October 720} 
\author{%
\textsc{Ois\'{\i}n A. P. Mac Conamhna}
\thanks{59 Farrer Road, Hornsey, London N8 8LD, United Kingdom. Email: macconamhna`at'hotmail.co.uk} 
\\[1ex] 
\normalsize {} 
}
\date{\today} 
\renewcommand{\maketitlehookd}{%
\begin{abstract}
{\noindent A historical record of a seismic tsunami is identified in the Irish annals for October 720 (all dates herein CE). It is contained in the earliest stratum of the annals, which survives in the form of a handful of iterated scribal copies of the foundational text of the tradition. This was compiled by the contemporary observation of noteworthy events for the years c. 563-740 at the monastery of Iona in the Scottish Hebrides. The 720 event is close outside the 2$\sigma$ radiocarbon {\it terminus ante quem} date ranges for tsunami deposits identified at Dury Voe (530-660) and Basta Voe (430-650) in the Shetland Isles, and is identified as a candidate progenitor. The possibility of the existence of associated tsunami deposits in Scotland or on the north coast of Ireland is highlighted.  }
\end{abstract}
}

\begin{document}
\maketitle


\section{Introduction}

Current understanding of the Irish annalistic tradition is that the compilation of contemporary observations of noteworthy events in chronological order was initiated in the Gaelic world at the monastery of Iona in the Scottish Hebrides, at or around its foundation in 563, \cite{Smyth}, \cite{McCarthy98}. This original text was maintained and extended there until around 740, when a copy of it was returned to Ireland, copied further, and extended locally thereafter in multiple locations. The first Gaelic annalist was in all probability Colm Cille, the founder of the Irish Christian missionary tradition, and a scribe. He established the monastery of Iona following his banishment from Ireland as penance for a battle fought over the copy of a book he had made, and one manuscript attributed traditionally to his hand - the Cathach, or Battler  - survives. The Annals of Iona are lost to Viking destruction, but consensus opinion is that the contemporary entries of the extant texts for c. 563-740 derive from them by repeated and independent scribal copy; and in at least one case, by copying and re-compiling at least three independent lineages of copies. Three of the extant  annals containing early material, the {\it Chronicon Scotorum} (\cite{SCOTORUM}), the Annals of Tigernach (\cite{TIGERNACH}), and the Annals of Ulster (\cite{ULADH}), are of relevance here. The {\it Chronicon Scotorum} records the following event for 720:

\begin{center}
Muirbrucht in mense Octimbris.\\
{\it A belch/bursting forth/huge tidal wave/eruption of the sea in the month of October}.
\end{center}

\noindent The Annals of Tigernach and of Ulster have the following for the same year, up to a minor spelling difference:

\begin{center}
Murbrucht mor in mense Octimbris.\\
{\it A great belch/bursting forth/huge tidal wave/eruption of the sea in the month of October}.\\
\end{center}

The given translations of {\it brucht} ({\it br\'ucht} in modern Irish spelling) are from \citet{ODonaill}; the meaning ``sudden rushing forth, any sudden burst or disruption'', \citet{Dwelly}, is retained in Scottish G\`{a}idhlig, the language derived from the Irish first brought to Scotland by the Ionan monks and their associated settlers. The Annals of Ulster then have the following unique record, not contained in any other extant Irish annal, as the second of nine entries for 721:

\begin{center}
Terrimotus in Octimbre.\\
{\it An earthquake in October}.\\
\end{center}

The term {\it terrimotus} was a generic one, literally ``earth-movement'', and could also be used to refer to landslides. It is translated here as ``earthquake'' because Iona lies north adjacent to the Great Glen Fault as it enters the sea on the west coast of Scotland, and experiences earthquakes. The chronology of all three texts is corrupt at this point. Read in isolation, the Annals of Ulster date the ``great bursting forth of the sea'' to 719, while the {\it Chronicon Scotorum} and Tigernach of themselves date it to 707. The given dates are according to the restored chronology of the texts, \cite{McCarthy98}, \cite{McCarthy05}.

\section{Discussion}
\noindent The common record in the three annals is clearly indicative of a tsunami; the only other slight interpretative possibility is of a major storm surge, but no storm is mentioned in conjunction with it; and the implied suddenness argues further against this interpretation. A tsunami large enough to leave a trace in the geological record is an extremely rare event in Scotland, with only three identified in the past 8,000 years, \cite{Bondevik05}; and the 720 record is the unique instance of its type in the annalistic corpus to the seventeenth century. Earthquakes are not uncommon in Scotland, but those large enough to leave an imprint in the annals were, on the evidence of the texts, infrequent. The prior probability of a tsunami and an independent significant earthquake happening one year apart and specifically in October, is thus vanishingly small. From a Bayesian perspective, the alternative possibility of a scribal error in dating one of them is much more probable. The texts themselves show clear evidence of the prevalence of dating errors, most obviously in the corruption of their own chronologies, and of most relevance here in the case of the Annals of Ulster, as is now explored.

While it was intrinsically robust, the distinctive calendrical system and its scribal notation used to define the original chronology of the Annals of Iona was baffling to the uninitiated, and liable to scribal error through momentary inattention or exemplar manuscript decay even in the hands of the expert. The passing of a year in the sequence of events was recorded by writing {\it K} or {\it Kl} to denote the event of the passing of the {\it Kalends} (first) of January. This was refined further originally by adding the {\it ferial}, a number from one to seven in Roman numerals, to denote the day of the week on which the {\it Kalends} of January occurred. {\it Kalends}-{\it ferial} pairs repeat periodically in the Julian calendar in a 28 year ``solar cycle". Used properly, this long repeating sequence gives the record of the chronology an internal robustness that allows errors to be identified and corrected. Between them, and them alone, the {\it Chronicon Scotorum} and the Annals of Tigernach preserve enough of it to allow a robust chronology of the earliest history of Ireland and Scotland to be reconstructed, \cite{McCarthy98}. However, the texts of themselves attest not only to scribal error, but also to the deliberate corruption of the chronology as written. The {\it feria} were not preserved at all by some texts, or in passages of others, and in those circumstances, the omission of the mark {\it Kl} from a block of text through scribal oversight (or otherwise) would lead to the annal losing a year from its chronology. 

The three surviving annals of interest manifestly share source material for the period that derives from a common, ``canonical'' source; and the common record of the ``bursting forth of the sea'' in 720 clearly derives from it. The canonical source is inferred to be the first copy of the Annals of Iona that returned to Ireland. A lost ``Annals of Bangor'', of that monastery in County Down on the northeastern coast of Ireland, has been proposed for this, given the Ulster focus of the early post-740 material, and the intimate historical connection between the monasteries of Bangor and Iona, \cite{Smyth}. The {\it Chronicon Scotorum} and the Annals of Tigernach collectively are understood to preserve the closest extant copy of this, including the original chronology. The Annals of Ulster are markedly different, in that while they draw on the canonical source, they also draw explicitly on other, apparently independent, lineages of copies of the Annals of Iona. They survive as a late fifteenth century manuscript in the hand of Ruaidhr\'{\i} \'{O} Luin\'{\i}n, that dates events according to the CE year in which they occurred, reflecting a conscious effort to convert away from the {\it Kalends}-{\it ferial} chronology, and to address some of the scribal errors and inconsistencies in it that had by then accumulated. The restoration of the chronology was not done perfectly, such that it still lagged by a year in 720. Due to the linguistic forms the Annals of Ulster contain, this re-dating could not have preceded the eleventh century, and it has been attributed to Cuan hua Lothcháin, who died in 1024, \cite{McCarthyUlster}. The Annals of Ulster cite at least three distinct sources: a canonical source, derived from the ``Annals of Bangor'' (cited obliquely in dating the death of Patrick to 492 as ``Dicunt Scoiti''/``The Irish say''); the lost Liber Cuanach, cited explicitly thirteen times, lastly for 629; and the form of words ``secundum alium librum'' (or the equivalent in a different word order: ``according to another book'', which could have been one or more other sources) is used eight times in the material to 1065. From what the Annals of Ulster have to say about the Liber Cuanach in particular it contained, at least in part, a copy of the Annals of Iona of an independent lineage to the canonical source. The chronology of these additional sources (presumably even when restored as well as they were able by the Ulster compiler) was inconsistent, at least at times, with the CE chronology adopted by the Annals of Ulster themselves, as the following entries make clear.\\

\noindent For 602 (dated by the text to 603):
\begin{center}
Omnia quae scripta sunt in anno subsequente, inueni in Libro Cuanach in isto esse perfecta.   
{\it Everything that has been set down in the following year, I have found in the Liber Cuanach to have taken place in this}.
\end{center}

The deaths in 665 of Diarmait, Blamac and Feichíne of the great mortality ``ie [the] Buidhe [Yellow of] Chonaill'' are dated by the text to 664, but it also records that they took place in 667 ``according to another book''.   Likewise, Guaire of Aidne died in 663, their death is recorded in the text both in 662 and in 665 ``according to another book''. In a unique entry, a lunar eclipse in 718 is dated by the text to the year 717. In truth the eclipse happened in 719; this implies that the Annals of Ulster drew the unique record of this eclipse from another source whose chronology was two years behind at that point. Finally, the death of Donnchad son of Brian B\'oramha is given to 1064 (1064 in the text, which by then had rectified its chronology) and also to 1065 ``according to another book''.

The clear evidence of these examples is that the chronology of the various sources of the Annals of Ulster was often inconsistent with its own, which was itself incorrect in 720; and that events could be mis-dated by several years either side of their true occurrence by these sources. As the earthquake record of 721 is unique to the Annals of Ulster, it clearly came from ``another book'', and its chronology is therefore particularly liable to be incorrect. The interpretation that it is from ``another book'' is re-enforced by the fact that an October entry appears as the second of nine entries for 721, suggesting chronological displacement within the events of that year, when events were recorded in strict chronological order by any individual original text. Furthermore, given medieval Irish knowledge of seismology, its essential contemporaneous relationship to the ``great bursting forth of the sea'' of 720 would not have been apparent to the Ulster compiler.   

The dating of events to the precision of a month or better is very uncommon in all of the texts. To have two events drawn from independent lineages dated specifically to the month of October one year apart is in and of itself sufficient grounds to infer their likely contemporaneity, and scribal error in the relative dating of one of them. Given the nature of the events and the other evidence discussed above, the conclusion that they are in fact contemporaneous appears to be overwhelmingly probable. They are therefore identified as the record of a seismic tsunami, observed at Iona in October 720. It is further proposed that the reason the record of it survived in the extant split form is as follows. It is proposed that the Annals of Iona originally recorded both the earthquake and the tsunami, but that only the record of the tsunami was copied into the canonical ``Annals of Bangor'', and so propagated to the {\it Chronicon Scotorum}, Tigernach and Ulster; while only the record of the earthquake was copied, with ultimately one year of relative chronological corruption, into an independent and now otherwise defunct lineage of copies, one of which was accessed by the compiler of the Annals of Ulster, who did not recognise its contemporaneity with the ``great bursting forth of the sea''.

\section{Conclusions}

A large seismic tsunami experienced in the Hebrides in 720 would be expected to leave a discernable signature in the geological record. A review of the literature of Atlantic tsunami deposits (\cite{Costa}) indicates one possibility, identified at Basta Voe and Dury Voe in the Shetland Isles. A thin layer of sand has been detected in littoral peat bogs at both locations. No potentially tsunamigenic submarine landslides of comparable age have been identified in adjacent basins. Radiocarbon dating of samples taken as close as possible above the sand layer gives a 2$\sigma$ radiocarbon {\it terminus ante quem} date range of [430,650] for the deposit at Basta Voe (\cite{Bondevik05}, \cite{Dawson06}),  and [530,660] at Dury Voe (\cite{Bondevik05}). Given the proximity of the date of the historical record to the upper limit of these {\it terminus ante quem} ranges, and the residual uncertainty in the calibration of littoral samples due to the marine reservoir effect, the 720 event is identified as a candidate progenitor of these deposits.  It is indicative of the tsunamigenic potential of the Great Glen Fault at that time. The Shetland Isles are also situated on this fault, 560km northeast of Iona on the other side of the Scottish mainland. Whether or not the Dury Voe and Basta Voe deposits may be better explained by the 720 event alone or by multiple significant earthquakes along the Great Glen Fault in that period, a seismic origin is proposed for their progenitor. The possibility of similar tsunami deposits dating to 720 in the Hebrides, on the west coast of Scotland, and on the north coast of Ireland, is highlighted for future fieldwork.

\section{Data Availability}
The input data to this communication were the Irish annalistic corpus in the form made available online by the Corpus of Electronic Texts (CELT), https://celt.ucc.ie, for the {\it Chronicon Scotorum}, the Annals of Tigernach, Ulster, Inisfallen, Connacht, Loch C\'e, the Kingdom of Ireland (Four Masters), and the Miscellaneous and Fragmentary Annals; as well as the Annals of Clonmacnoise, \cite{CLONMACNOISE}, Roscrea, \cite{ROSCREA}, and Boyle, \cite{BOYLE}; along with the other cited works. No further data beyond this communication were generated.

\bibliography{Tsunami_In_Irish_Annals_final}

\begin{thebibliography}{15}
\providecommand{\natexlab}[1]{#1}
\providecommand{\url}[1]{\texttt{#1}}
\expandafter\ifx\csname urlstyle\endcsname\relax
  \providecommand{\doi}[1]{doi: #1}\else
  \providecommand{\doi}{doi: \begingroup \urlstyle{rm}\Url}\fi

\bibitem[Bondevik et~al.(2005)Bondevik, Mangerud, Dawson, Dawson, and
  Lohne]{Bondevik05}
S.~Bondevik, J.~Mangerud, S.~Dawson, A.~Dawson, and {\O}.~Lohne.
\newblock {Evidence for three North Sea tsunamis at the Shetland Islands
  between 8000 and 1500 years ago}.
\newblock \emph{Quaternary Science Reviews}, 24:\penalty0 1757--1775, 2005.

\bibitem[Costa et~al.(2021)Costa, Dawson, Ramalho, Engel, Dourado, Bosnic, and
  Andrade]{Costa}
P.~J.~M. Costa, S.~Dawson, R.~S. Ramalho, M.~Engel, F.~Dourado, I.~Bosnic, and
  C.~Andrade.
\newblock {A review of onshore tsunami deposits along the Atlantic coasts}.
\newblock \emph{Earth-Science Reviews}, 212:\penalty0 103441, 2021.

\bibitem[{D. Gleeson and S. MacAirt}(1959)]{ROSCREA}
{D. Gleeson and S. MacAirt}.
\newblock {The annals of Roscrea}.
\newblock \emph{Proceedings of the Royal Irish Academy}, 59C:\penalty0 137--80,
  1959.

\bibitem[Dawson et~al.(2006)Dawson, Dawson, and Bondevik]{Dawson06}
A.~G. Dawson, S.~Dawson, and S.~Bondevik.
\newblock {A late Holocene Tsunami at Basta Voe, Yell, Shetland Isles}.
\newblock \emph{Scottish Geographical Journal}, 122:\penalty0 100--108, 2006.

\bibitem[Dwelly(1994)]{Dwelly}
E.~Dwelly.
\newblock \emph{{Faclair G\`{a}idhlig gu Beurla}}.
\newblock Gairm Publications, 1994.

\bibitem[Freeman(1924-7)]{BOYLE}
A.~Freeman.
\newblock {The annals in Cotton MS Titus A xxv [Annals of Boyle]}.
\newblock \emph{Revue Celtique}, 41-4:\penalty0 301--30, 283--305, 358--84,
  336--61, 1924-7.

\bibitem[McCarthy(1998)]{McCarthy98}
D.~P. McCarthy.
\newblock {The Chronology of the Irish Annals}.
\newblock \emph{Proceedings of the Royal Irish Academy: Archaeology, Culture,
  History, Literature}, 98C:\penalty0 203--255, 1998.

\bibitem[McCarthy(2002)]{McCarthyUlster}
D.~P. McCarthy.
\newblock {The chronological apparatus of the {\it Annals of Ulster}, AD
  82-1019}.
\newblock \emph{Peritia}, 16:\penalty0 256--283, 2002.

\bibitem[McCarthy(2005)]{McCarthy05}
D.~P. McCarthy.
\newblock {Chronological Synchronisation of the Irish Annals, Fourth Edition}.
\newblock \emph{{Online resource,
  https://www.scss.tcd.ie/misc/kronos/chronology/synchronisms/annals-chron.htm,
  accessed 05/11/2022}}, 2005.

\bibitem[Murphy(1896)]{CLONMACNOISE}
D.~Murphy.
\newblock \emph{{The Annals of Clonmacnoise}}.
\newblock Dublin, 1896.

\bibitem[\'{O}~D\'{o}naill(1977)]{ODonaill}
N.~\'{O}~D\'{o}naill.
\newblock \emph{{Focl\'{o}ir Gaeilge-B\'{e}arla (2012 edition)}}.
\newblock An G\'{u}m, 1977.

\bibitem[Smyth(1972)]{Smyth}
A.~P. Smyth.
\newblock {The Earliest Irish Annals: Their First Cotemporary Entries, and the
  Earliest Centres of Recording}.
\newblock \emph{Proceedings of the Royal Irish Academy: Archaeology, Culture,
  History, Literature}, 72:\penalty0 1--48, 1972.

\bibitem[{Unknown}(353--1150)]{SCOTORUM}
{Unknown}.
\newblock {The Chronicon Scotorum}.
\newblock \emph{{Corpus of Electronic Texts
  (CELT),https://celt.ucc.ie/published/G100016/index.html, accessed
  05/11/2022}}, 353--1150.

\bibitem[{Unknown}(431--1541)]{ULADH}
{Unknown}.
\newblock {The Annals of Ulster}.
\newblock \emph{{Corpus of Electronic Texts
  (CELT),https://celt.ucc.ie/published/G100001A/index.html, accessed
  05/11/2022}}, 431--1541.

\bibitem[{Unknown}(488--1178)]{TIGERNACH}
{Unknown}.
\newblock {The Annals of Tigernach}.
\newblock \emph{{Corpus of Electronic Texts (CELT),
  https://celt.ucc.ie/published/G100002/index.html, accessed 05/11/2022}},
  488--1178.

\end{thebibliography}


\end{document}